\documentclass[11pt]{article}
\usepackage{blois,epsfig}

\def\be{\begin{equation}}
\def\ee{\end{equation}}
\def\bea{\begin{eqnarray}}
\def\eea{\end{eqnarray}}

\begin{document}
\begin{flushright}
TUM-HEP-780/10\\
\end{flushright}
\vspace*{4cm}
\title{Highlights of Flavour Violation in the Presence of a 4th Generation}

\author{Tillmann Heidsieck}

\address{Physik Department, Technische Universit\"at M\"unchen,
James-Franck-Stra{\ss}e, \\D-85748 Garching, Germany}

\maketitle\abstracts{
We present a short overview of recent work on the topic of flavour violation in
the presence of a sequential fourth generation (4G) of quarks and leptons. We discuss the implications
of this simple extension of the Standard Model (SM3) for rare decays and CP violating observables.
The observables of interest were chosen with the start of the LHC, the approval of SuperBelle and
the upcoming NA62 experiment in mind. A special emphasis is placed on the importance of $\varepsilon^\prime/\varepsilon$ as
a constraint for this model.
}

\section{Introduction}
The addition of a fourth generation to the Standard Model (hereafter referred to by SM4) is one of the most simple extensions of the SM3.
While the setup usually studied (perturbative Youkawa couplings) does not explain any of the known theoretical shortcomings present in the SM3,
the addition of a 4G could explain many current tensions between flavour observables like $S_{\psi\phi}$, $S_{\phi K_S}$,
the di-muon charge asymmetry measured at D0 or even the slight tension between $\varepsilon_K$ and $S_{\psi K_S}$.
In the presence of a 4G electroweak baryogenesis might be viable \cite{Hou:2008xd}. In models with non-pert. Youkawa couplings
electroweak symmetry breaking might be triggered by the heavy fourth generation quarks\cite{Holdom:1986rn}.
Over the recent years the 4G has attracted renewed attention\cite{Holdom:2009rf} with studies of EWP data\cite{Eberhardt:2010bm,Chanowitz:2009mz,Erler:2010sk} 
and flavour observables \cite{Bobrowski:2009ng,Hou:2005yb,Hou:2006mx,Arhrib:2006pm,Soni:2008bc,Soni:2010xh,Buras:2010pi,Buras:2010nd,Buras:2010cp}.
This is a testament to the fact that though the 4G only introduces a very small number of additional parameters it can not be
ruled out, yet. The generalisation of the CKM matrix to four generations yields five new parameters
\begin{equation}
 \theta_{14},\theta_{24},\theta_{34},\delta_{14},\delta_{24}\,.
\end{equation}
Together with the two new quark masses this gives a total of $7$ new parameters. 
It was however realised that there are severe constrains on these parameters coming from from CKM unitarity, 
oblique corrections\cite{Eberhardt:2010bm,Chanowitz:2009mz,Erler:2010sk} and FCNC processes\cite{Bobrowski:2009ng,Soni:2010xh,Buras:2010pi}
which can be summarised by the following approximate bounds
\begin{eqnarray}
 &&s_{14} \le 0.04\,,\quad  s_{24} \le 0.17\,,\quad    s_{34} \le 0.27\,,\\
 &&300{\rm GeV}\le m_{t^\prime} \le {\rm Min}(600{\rm GeV}, M_W/|s_{34}|)\,.
\end{eqnarray}
The second line in the above reflects the important contribution of the mixing between third and fourth generation to the $T$ parameter\cite{Eberhardt:2010bm,Chanowitz:2009mz}.\\

In presenting the results of the global analysis, it will be useful to
use  a special colour coding,  in order to emphasise
some aspects of the anatomy presented
in the next section and to stress certain points that we found in the process
of our numerical analysis:
\begin{itemize}
\item  The large black point represents the SM3.
\item Light blue and dark blue  points stand for the results of our global analysis of the SM4 
with the following distinction: light blue stands for  ${\rm Br}(K_L\rightarrow \pi^0\nu\bar\nu) > 2\cdot 10^{-10}$
and dark blue for ${\rm Br}(K_L\rightarrow \pi^0\nu\bar\nu) \leq 2\cdot 10^{-10}$. 
Note that the regions with light and dark blue points are
not always exclusive but that the dark blue points are plotted above the
light blue ones.
\item The yellow, green and red colours represent
 the three scenarios  for $S_{\psi\phi}$ and ${\rm Br}(B_s\to\mu^+\mu^-)$ 
that are shown in Table~\ref{tab:Bscenarios}.
\end{itemize}

\begin{table}[t!!!pb]
\begin{center}
\begin{tabular}{|c||c|c|c|}
\hline
 		&BS1 (yellow) &BS2 (green)	& BS3 (red) 	\\ \hline\hline
$S_{\psi\phi}$	& $0.04\pm 0.01$& $0.04\pm 0.01$ & $ \geq 0.4$ 	\\ \hline
${\rm Br}(B_s\to\mu^+\mu^-)$	& $(2\pm 0.2)\cdot 10^{-9}$ & $(3.2\pm 0.2)\cdot 10^{-9} $   &   $\geq 6\cdot 10^{-9}$ 	\\ \hline
\end{tabular}
\caption{Three scenarios for  $S_{\psi\phi}$ and  ${\rm Br}(B_s\to\mu^+\mu^-)$.} \label{tab:Bscenarios}
\end{center}
\end{table}

\section{Rare Decays and CP violation}
\subsection{General predictions}
With the start of the LHC earlier this year, the prospects of measuring the rare decays $B_q\rightarrow \mu^+\mu^-$ in the next years are very promising. Together with the long standing hint
of additional CP violation in $B_s$ decays this motivates the study of the correlation between ${\rm Br}(B_q\rightarrow \mu^+\mu^-)$ and $S_{\psi\phi}$.
\begin{figure}[ht] 
\begin{center}
\includegraphics[width=.48\textwidth]{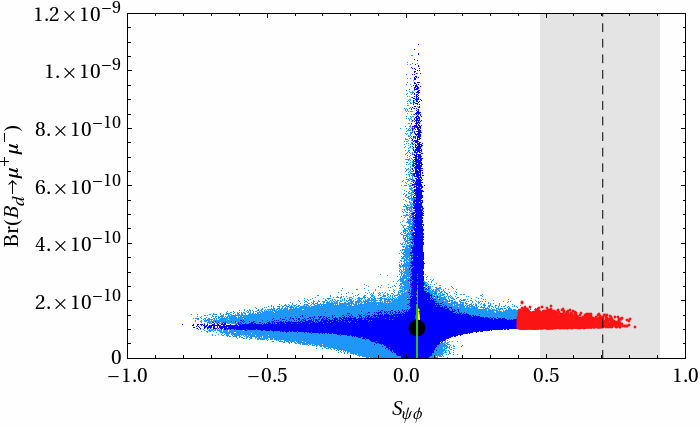}\hspace{.03\textwidth}
\includegraphics[width=.48\textwidth]{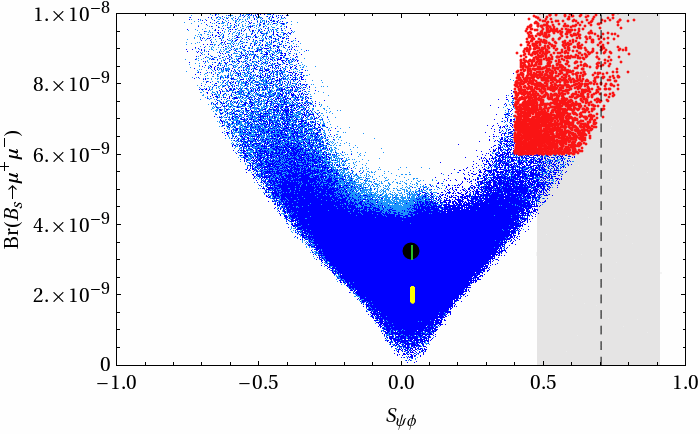}
\end{center}
\vspace{-.5cm}
\caption{${\rm Br}(B_d\to\mu^+\mu^-)$ (left panel) and ${\rm Br}(B_s\to\mu^+\mu^-)$ (right panel) each as a function of $S_{\psi\phi}$. \label{fig:Spsiphi-Bmumu}}
\end{figure}
In Fig.~\ref{fig:Spsiphi-Bmumu} we show on the left panel ${\rm Br}(B_d\to\mu^+\mu^-)$ and on the right panel ${\rm Br}(B_s\to\mu^+\mu^-)$ each as a function of $S_{\psi\phi}$.
The first striking feature is the very different structure of those two correlations, this is a manifestation of the non-(C)MFV structure of the SM4. While both branching ratios
can be enhanced (or suppressed) individually by roughly one order of magnitude a simultaneous enhancement (or suppression) is not possible. Furthermore an enhanced $S_{\psi\phi}$
would force ${\rm Br}(B_d\to\mu^+\mu^-)$ to be SM3 like while possibly enhancing ${\rm Br}(B_s\to\mu^+\mu^-)$. The measurement of ${\rm Br}(B_s\to\mu^+\mu^-)$ would also restrict
the possible values for the branching ratios ${\rm Br}(B\to X_s \gamma)$ and ${\rm Br}(B\to X_s \ell^+\ell^-)$ and might therefore help rule out this model.\\[2mm]

\begin{figure}[t!!!pb] 
\begin{center}
\includegraphics[width=.48\textwidth]{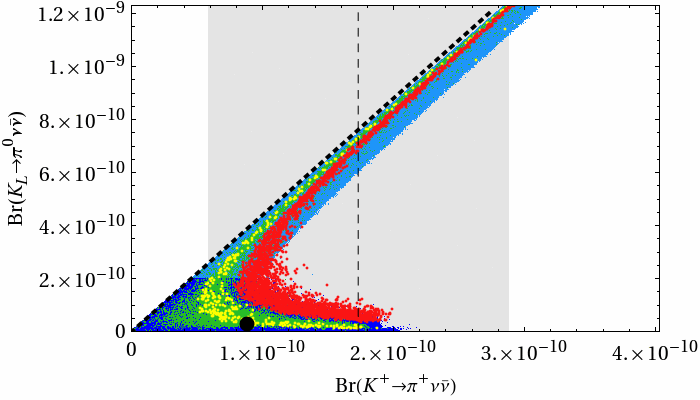}
\end{center}
\vspace{-.5cm}
\caption{${\rm Br}(K_L\to\pi^0\nu\bar\nu)$ as a function of ${\rm Br}(K^+\to\pi^+\nu\bar\nu)$. The dotted line corresponds to the model-independent GN bound.\label{fig:Kpinunu}}
\end{figure}

Among the many decay modes of the Kaon, the rare decays $K_L\to \pi^0\nu\bar\nu$ and $K^+\to \pi^+\nu\bar\nu$ are among the theoretically cleanest. In Fig.~\ref{fig:Kpinunu} we
show the correlation between the branching ratios of $K_L\to \pi^0\nu\bar\nu$ and $K^+\to \pi^+\nu\bar\nu$.
Clearly an enhancement by orders of magnitude for ${\rm Br}(K_L\to\pi^0\nu\bar\nu)$ is still possible while we observe only a mild correlation with the $B$ system. Note that an enhancement
of ${\rm Br}(K_L\to\pi^0\nu\bar\nu)$ by more than one order of magnitude would imply an enhancement of ${\rm Br}(K^+\to\pi^+\nu\bar\nu)$ while the reverse is not true. The cut of  ${\rm Br}(K^+\to\pi^+\nu\bar\nu) \le 2.2\cdot 10^{-10}$ on the lower branch
is due to the constraint from ${\rm Br}(K_L\to \mu^+\mu^-)_{\rm SD}\le 0.8\cdot 10^{-9}$.

\boldmath
\subsection{The importance of $\varepsilon^\prime/\varepsilon$}
\unboldmath
The ratio of direct over indirect CP violation in the $K$ system usually referred to as $\varepsilon^\prime/\varepsilon$ can in principle pose very stringent bounds on many models beyond the SM3. However due to 
theoretical difficulties with non-perturbative hadronic matrix elements this constraint has to be used with great care. We choose to study different scenarios for the non-pert. parameters in order to understand
the constraint $\varepsilon^\prime/\varepsilon$ can pose for the SM4. In the following the colour-code of the figures will be as described in Tab.~\ref{tab:Rscenarios}.\\[2mm]
\begin{table}[ht] 
\begin{center}
\begin{tabular}{|c|c||c|}
\hline
$R_6$	& $R_8$	&	 	\\ \hline\hline
$1.0$	& $1.0$	& dark blue	\\ \hline
$1.5$	& $0.8$	& purple	\\ \hline
$2.0$	& $1.0$	& green		\\ \hline
$1.5$	& $0.5$	& orange	\\ \hline
\end{tabular}
\caption{Four scenarios for the parameters $R_6$ and $R_8$\label{tab:Rscenarios}}
\end{center}
\end{table}

\begin{figure}[ht]
\begin{center}
\includegraphics[width=.48\textwidth]{fig5-2.png}\hspace{.03\textwidth}
\includegraphics[width=.48\textwidth]{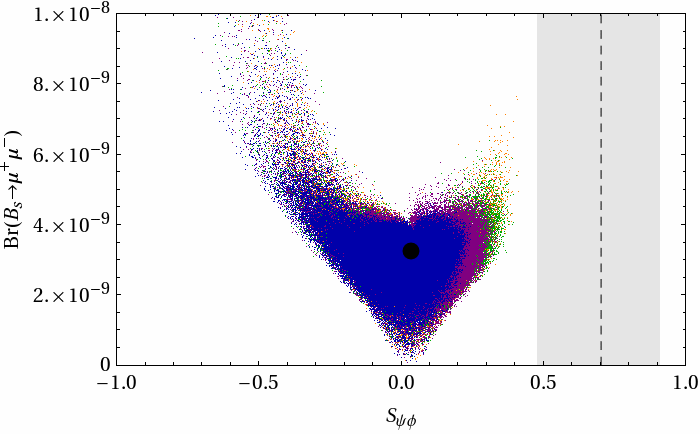}
\caption{The correlation ${\rm Br}(B_s\to\mu^+\mu^-)$ vs. $S_{\psi\phi}$ before and after including the $\varepsilon^\prime/\varepsilon$-constraint (colour-coding according to Tab.~\ref{tab:Rscenarios}).\label{fig:corr-eps1}}
\end{center}
\end{figure}

In Fig.~\ref{fig:corr-eps1} we show the impact of $\varepsilon^\prime/\varepsilon$ on the correlation between ${\rm Br}(B_s\to\mu^+\mu^-)$ and $S_{\psi\phi}$ introduced earlier. The most striking feature here is the asymmetrical
nature of the constraint on this correlation.
For $S_{\psi\phi}>0$ the contributions of the SM3 and the additional $t^\prime$ contribution have the same sign, therefore the $Z$ penguins with $t$ and $t^\prime$ overcompensate the QCD penguins thereby pushing $\varepsilon^\prime/\varepsilon$
far below the observed value. If one decreases $B_8$ and increases $B_6$ the influence of the $Z$ penguins can be lessened while strengthening the QCD penguins, which partly circumvents the problem (orange points).

\section{Conclusions}
The main results of our analysis\cite{Buras:2010cp} can be summarised as
\begin{itemize}
 \item Many of the observed tensions in the flavour sector can be resolved simultaneous in the SM4.
 \item The branching ratio ${\rm Br}(B_s\to\mu^+\mu^-)$ can be enhanced or suppressed in the SM4. However if $S_{\psi\phi} \gg 0$ as suggested by the Tevatron data was indeed true we would expect an enhancement of ${\rm Br}(B_s\to\mu^+\mu^-)$.
 \item In the $K$ system there is independently of the $B$ system much room for in some cases huge effects, however they are correlated among each other.
 \item $\varepsilon^\prime/\varepsilon$ can pose a very stringent constraint on the SM4 if the non-pert. parameters $B_6$ and $B_8$ were known to a decent accuracy.
\end{itemize}

\section*{Acknowledgements}
I want to thank Andrzej J. Buras, Bj\"orn Duling, Christoph Promberger, Thorsten Feldmann and Stefan Recksiegel for the fruitful collaboration aswell as  P.Q. Hung for the invitation to this great 21th Rencontres des Blois.
This work was partially supported by GRK 1054 of Deutsche Forschungsgemeinschaft.

\end{document}